Title
**Experimental long-distance quantum secure direct communication**

The authors
**Feng Zhu**,
Tsinghua National Laboratory for Information Science and Technology, Department of Electronic Engineering, Tsinghua University, Beijing, 100084, China
fzhu11@tsinghua.edu.cn

***Wei Zhang**,
Tsinghua National Laboratory for Information Science and Technology, Department of Electronic Engineering, Tsinghua University, Beijing, 100084, China
zwei@tsinghua.edu.cn

**Yubo Sheng**,
College of Telecommunications and Information Engineering, Nanjing University of Posts and Telecommunication, Nanjing, 210003, China
shengyb@njupt.edu.cn

**Yidong Huang**,
Tsinghua National Laboratory for Information Science and Technology, Department of Electronic Engineering, Tsinghua University, Beijing, 100084, China
yidonghuang@tsinghua.edu.cn

The information of the corresponding author
**Prof. Wei Zhang**
Tsinghua National Laboratory for Information Science and Technology
Department of Electronic Engineering, Tsinghua University
Address: Room 2-101, Rohm Building, Tsinghua University, Beijing 100084, China
Tel: +86-10-6279-7073-802
Fax: +86-10-6279-7073-807
E-mail: zwei@tsinghua.edu.cn

# Experimental long-distance quantum secure direct communication

**Feng Zhu[1]**, **Wei Zhang*[1]**, **Yubo Sheng[2]**, **Yidong Huang[1]**

1. Tsinghua National Laboratory for Information Science and Technology, Department of Electronic Engineering, Tsinghua University, Beijing, 100084, China

2. College of Telecommunications and Information Engineering, Nanjing University of Posts and Telecommunication, Nanjing, 210003, China

   * *Corresponding author: Wei Zhang, [zwei@tsinghua.edu.cn](zwei@tsinghua.edu.cn)*

**Abstract**

Quantum secure direct communication (QSDC) is an important quantum communication branch, which realizes the secure information transmission directly without encryption and decryption processes. Recently, two table-top experiments have demonstrated the principle of QSDC. Here, we report the first long-distance QSDC experiment, including the security test, information encoding, fiber transmission and decoding. After the fiber transmission of 0.5 km, quantum state fidelities of the two polarization entangled Bell states are 91% and 88%, respectively, which are used for information coding. We theoretically analyze the performance of the QSDC system based on current optical communication technologies, showing that QSDC over fiber links of several tens kilometers could be expected. It demonstrates the potential of long-distance QSDC and supports its future applications on quantum communication networks.

**Keywords**

Quantum secure direct communication, polarization entangled Bell state, optical fibers, Bell's inequality test, Bell state measurement

**Introduction**

Utilizing principles of quantum mechanics to realize new functions beyond classical communication, quantum communication is an important field in quantum information science and technologies [1-3]. Optical fiber is a widely used quantum channel to demonstrate long-distance quantum communication applications due to low attenuation and weak decoherence when photons propagate in it. A plenty of quantum communication protocols have been experimentally investigated in fiber channels, such as quantum teleportation [4-6], quantum secret sharing [7] and quantum key distribution [8-10], showing their great potential in future quantum communication networks [11,12].

As an important branch of quantum communication, quantum secure direct communication (QSDC) [13-15] attracts much attention in recent years. On the conceptual side, the first QSDC protocol exploits quantum entanglement [13] and the standard criterion for QSDC was explicitly clarified subsequently [14]. Furthermore, single photons can also realize QSDC, which is feasible in current single-photon devices [15]. In these protocols, the security of quantum channel is tested firstly while the photons are stored in the memory. Then information is encoded on the photons, transferred over quantum channel and decoded by quantum state discrimination. Since the information is transferred without encryption and decryption in QSDC protocols, the security loophole in key managements is avoided [13]. Furthermore, QSDC also can be used to transfer the key for cryptography securely, which has the potential for higher capacity than quantum key distribution [16]. On the experimental side, two table-top experiments were reported recently for proof-of-principle demonstrations of QSDC protocols based on single photons and entanglement [17,18]. In the entanglement-based QSDC experiment, the photon pairs are transferred in free-space and stored in quantum memories based on atomic ensembles [17]. The information is encoded on polarizations of entangled photon pairs and discriminated by quantum state tomography (QST) technology [19]. In the single-photon based QSDC experiment, the photons are transferred in optical fiber and stored in a fiber coil, the information is transferred against the noise and loss by frequency coding [18].

In this paper, we report the first long-distance entanglement-based QSDC experiment. The entangled photon pairs are stored in fiber coils, which have higher efficiency than atomic quantum memories at telecom band[17]. After security test of the quantum channel, the information is encoded on polarizations of the photon pairs. Then the photons are transferred over optical fibers of 0.5 km. In the process of information decoding, we use the Bell state measurement (BSM) to take over the density matrix constructing [17], showing that the fidelities of the two polarization entangled Bell states are 91% and 88%, respectively, after their fiber transmission. We analyze the performance of the QSDC system theoretically and shows that QSDC over a fiber link of several tens kilometers could be expected under current optical communication technologies. It provides a way to extend QSDC to longer distance, which would support its future applications on quantum communication networks.

**Materials and methods**

**A. Entanglement-based QSDC protocol**

The entanglement-based QSDC protocol investigated in this work has following steps [14].

Step 1. Alice produces photon pairs which are in the polarization entangled Bell state of $|\psi^+\rangle=(|H\rangle_A|V\rangle_B+|V\rangle_A|H\rangle_B)/\sqrt{2}$, where 'H' and 'V' denote the horizontal and vertical polarizations, respectively.

Step 2. Alice sends one of the photons in a pair to Bob, which has the subscript of 'B', through a piece of fiber 500 meters in length in this experiment. The other photon is kept by Alice and indicated by the subscript of 'A'.

Step 3. Bob receives the photons 'B'. He extracts a part of them by a switch and measures them. Then, he sends the results to Alice. Rest of the photons "B" are stored in a quantum memory. Alice also extracts a part of photons 'A' by another switch and measures them. Combining with the measurement results of Bob, she realizes the Bell's inequality test of the polarization entangled Bell state to ensure the security of the quantum channel. Rest of the photons "A" are also stored in a quantum memory.

Step 4. If the security test shows that the quantum channel is secure, Alice encodes information on the polarization entangled state of the photon pairs by manipulating the polarization of rest photons of 'A', which output from the quantum memory at Alice side. For example, '1' is encoded by $|\psi^-\rangle=(|H\rangle_A|V\rangle_B-|V\rangle_A|H\rangle_B)/\sqrt{2}$ and '0' is encoded by $|\psi^+\rangle=(|H\rangle_A|V\rangle_B+|V\rangle_A|H\rangle_B)/\sqrt{2}$. Then Alice sends the photons 'A' to Bob.

Step 5. Bob receives the encoded photons "A" and takes a BSM between them and the photons 'B' output from the quantum memory at Bob side. The BSM decodes the information carried by the photon pair.

**B. Experimental setup**

The experimental setup is shown in Fig. 1. At Alice side, telecom-band frequency-degenerated polarization entangled photon pairs are generated by a quantum light source based on spontaneous four wave mixing in optical fibers [20]. Their biphoton state is $|\psi^+\rangle = (|H\rangle_A |V\rangle_B + |V\rangle_A |H\rangle_B)/\sqrt{2}$. The two photons in a pair are denoted by 'A' and 'B', which output from Port A and Port B, respectively. Both photons are at 1549.32 nm. In the experiment, the pump level of the source is controlled to realize a coincidence to accidental coincidence ratio (CAR) of 50 when the pairs are measured by the coincidence counting system directly. Under this condition, the visibilities of the two photon interferences under two non-orthogonal polarization bases (0 degree and 45 degree) are 96% and 90%, respectively, showing the property of polarization entanglement. The details about the source and the coincidence counting system are presented in the Supplementary Materials.

The photons 'B' are sent to Bob though a piece of dispersion shifted fiber (DSF) 500 meters in length, then distributed into two paths by a 50:50 fiber coupler. In one path, half of the photons are measured by a single photon detector (SPD-2) after a half wave plate (HWP-2) and a polarization beam splitter (PBS-2). The detection records are sent back to Alice by a public channel. At Alice side, the photons 'A' are also distributed into two paths by a 50:50 fiber coupler. Half of them are detected by a single photon detector (SPD-1) after another set of half wave plate and polarization beam splitter (HWP-2, PBS-2). By proper polarization aliments and polarization base selections for the detections (through the adjustment of HWP-1 and HWP-2 and the port selection of PBS-1 and PBS-2), the Bell's inequality is tested by the coincidence counting of the detection records of SPD-1 and SPD-2. The result of Bell's inequality test is used to check the security when the photons 'B' are sent to Bob.

At Alice side, the photons 'A' at the other path are sent to a piece of DSF, which is used as the quantum memory (QM-1). The memory is used to wait the result of security test. In the experiment, the fiber length of QM-1 is 2200 meters, providing a memory time of 11 μs. Then, the information is encoded on photons 'A' by two quarter wave plates (QWP-1 and QWP-2) when they output from

QM-1. If the optical axes of the two QWPs are all aligned to "H", the biphoton state of the photon pairs changes to $|\psi^-\rangle=(|H\rangle_A|V\rangle_B-|V\rangle_A|H\rangle_B)/\sqrt{2}$, which is used to represent '1'. If the optical axes of the two QWPs are orthogonal, the biphoton state of the photon pairs is unchanged, which is $|\psi^+\rangle=(|H\rangle_A|V\rangle_B+|V\rangle_A|H\rangle_B)/\sqrt{2}$. It is used to represent '0'. After the encoding process, this part of photons 'A' are sent to Bob by another piece of DSF, which is also 500 meters in length.

At Bob side, the photons 'B' at the other path are also sent to a quantum memory (QM-2), which is a piece of DSF, waiting for the arrival of the encoded photons 'A'. The storage time of QM-2 is also 11 µs. These photons 'B' output from the QM-2 when the encoded photons 'A' arrive at Bob. Then they are sent into a polarization BSM system to discriminate their states. The main part of the BSM system includes a 50:50 fiber coupler and two polarization beam splitters (PBS-5 and PBS-6). The output ports of PBS-5 and PBS-6 are denoted by *a*, *b*, *c* and *d*, respectively, with polarization maintaining fiber (PMF) pigtails. Among the four ports, *a* and *c* are the ports for the photons with horizontal polarization (H), *b* and *d* are the ports for the photons with vertical polarization (V). Three polarization controllers (PC-1, PC-2 and PC-3) are used to calibrate the polarization bases in the setup. An optical delay line (ODL) is placed at one input port of the fiber coupler to make the two photons in a pair pass through the fiber coupler simultaneously. Then, the photon pairs are detected by two single photon detectors and the time-correlated single photon counting (TCSPC) module through two polarization beam splitters (PBS-3 and PBS-4). PBS-3 and PBS-4 are used to select the basis of the projection measurement in the BSM. Their ports are denoted by *e*, *f*, *g* and *h*, respectively, also with PMF pigtails. When the ports of *a*, *b*, *c* and *d* are connected to the ports of *e*, *f*, *g* and *h*, respectively, the coincidence counts of the setup record the photon pairs that one photon outputs from ports *a* or *b*, and the other outputs from *c* or *d*. This setting is indicated by "**M₁**". When the ports of *a*, *b*, *c* and *d* are connected to the ports of *e*, *g*, *f* and *h*, respectively, the coincidence counts of the setup record the photon pairs that one photon outputs from the port *a* or *c*, and the other outputs from the port *b* or *d*. This setting is indicated by "**M₂**".

**Results**

In the experiment, the function of the security test and the information transmission of the entanglement-based QSDC protocol are demonstrated separately. To show the security test in the setup, the Bell's inequality test is realized by the coincidence measurements of SPD-1 and SPD-2. The experiment result is shown in Table 1. $\theta_1$ and $\theta_2$ are angles of the optical axes of HWP-1 and HWP-2 to the vertical direction, respectively. Under each setting of HWP-1 and HWP-2, four coincidence counts are recorded, which are denoted by $N_{i,k}(\theta_1, \theta_2)$, $N_{j,l}(\theta_1, \theta_2)$, $N_{i,l}(\theta_1, \theta_2)$ and $N_{j,k}(\theta_1, \theta_2)$, respectively. Here, $i$ and $j$ denote the two output ports of PBS-1 and $k$ and $l$ denote the two output ports of PBS-2. Different basis for the projective measurement in the Bell's inequality test is realized by corresponding output ports combination. For each coincidence measurement, the counting time is 60 s.

The result of the Bell's inequality is calculated by [21]

$$S = E(\theta_1=0, \theta_2=11.25) + E(\theta_1=22.5, \theta_2=11.25) + E(\theta_1=22.5, \theta_2=33.75) - E(\theta_1=0, \theta_2=33.75)$$

Where, the expectation $E(\theta_1, \theta_2)$ is calculated by

$$E(\theta_1,\theta_2) = \frac{N_{i,k}(\theta_1,\theta_2)+N_{j,l}(\theta_1,\theta_2)-N_{i,l}(\theta_1,\theta_2)-N_{j,k}(\theta_1,\theta_2)}{N_{i,k}(\theta_1,\theta_2)+N_{j,l}(\theta_1,\theta_2)+N_{i,l}(\theta_1,\theta_2)+N_{j,k}(\theta_1,\theta_2)} \quad (1)$$

The result of S is 2.46±0.12, in which the error includes the contributions of counting statistics, angular position of HWPs and imperfection of the PBSs. It is larger than 2, violating the Bell's inequality by 4 standard deviations. It shows the security of the quantum channel.

On the other hand, the function of information transmission is demonstrated by another path at Alice and Bob. To show the photon pair source and the BSM system can realize all the functions, we set the biphoton state of the generated pairs to $|\psi^+\rangle$ and $|\psi^-\rangle$ through adjusting QWP-1 and QWP-2, by which the encoding process is simulated.

Firstly, we measure the photon pairs by the BSM system directly without the DSFs for transmission and quantum memories. For a specific state, the coincidence counts of SPD-3 and SPD-4 are measured under different time delay, which is adjusted by the ODL. The efficiencies of the two SPDs are set as 15% in the experiment. The coincidence counts are recorded in a time bin of 1.32ns covering the coincidence peak. The time for each measurement is 120 seconds. The experiment results are shown in Fig.2. Fig. 2a shows the results when the BSM system is set as "$\mathbf{M_1}$" to discriminate the state $|\psi^-\rangle$ from $|\psi^+\rangle$. The hollow squares and circles show the coincidence counts under different delay time of the ODL when the photon pairs are in $|\psi^-\rangle$ and $|\psi^+\rangle$, respectively. The red and green curves are their fitting curves using Gaussian functions. It can be seen that the results of $|\psi^-\rangle$ show a coincidence peak at the delay time of 466ps, which is the time that the two photons arrive the fiber coupler simultaneously. At this delay time, the result of $|\psi^+\rangle$ shows a dip. The visibility of the two fringe is 87%, which is calculated by

$$V = \frac{R_{max} - R_{min}}{R_{max} + R_{min}} \quad (2)$$

Where the visibility is denoted by $V$, $R_{max}$ and $R_{min}$ are the values of the peak and dip at the two fitting curves. It shows that the BSM system can discriminate $|\psi^-\rangle$ from $|\psi^+\rangle$ by the setting of "$M_1$". Fig. 2b shows the results that when the BSM system is set as "$M_2$" to discriminate the state $|\psi^+\rangle$ from $|\psi^-\rangle$. The hollow squares and circles also show the results for $|\psi^-\rangle$ and $|\psi^+\rangle$, respectively, with the red and green curves as their Gaussian fitting curves. It shows that in this case the result of $|\psi^+\rangle$ have a coincidence peak and those for $|\psi^-\rangle$ have a dip when the two photons arrive the fiber coupler simultaneously. The visibility of the peak and dip in the curves is 89% calculated by Eq. (2). It shows that the BSM system can discriminate $|\psi^+\rangle$ from $|\psi^-\rangle$ by the setting of "$M_2$". According to the results showing in Fig. 2, the fidelities of $|\psi^-\rangle$ and $|\psi^+\rangle$ are 95%±3% and 93%±3%, respectively. It demonstrates that the function of information encoding and decoding on the polarization entangled Bell states can be realized by the photon pair source and the BSM system in the experimental setup. It is worth noting that two measurements of "$M_1$" and "$M_2$" are used in the experiment since we only have two SPDs for the BSM. If there are four SPDs located at the port $a$, $b$, $c$ and $d$, the two states can be discriminated directly.

Then, we add the optical fibers for transmission and quantum memories to simulate the information transmission in the QSDC protocol. There are two pieces of DSFs connecting Alice and Bob for the two photons in a pair, respectively. Each of them has a length of 500 meters. On the other hand, two pieces of DSFs of 2.2 kilometers are placed in Alice and Bob sides as the quantum memories. Hence, the fiber length for the both photons is 2.7 kilometers before they are measured by the BSM system. The results are shown in Fig. 3, with similar experimental conditions with those of Fig. 2, except that the time of each coincidence measurement is 180 seconds. Fig. 3a and b show the results when the BSM system is set as "$M_1$" and "$M_2$", respectively. In both figures the hollow squares and circles are the coincidence counts when the photon pairs are in $|\psi^-\rangle$ and $|\psi^+\rangle$, respectively, with the red and green curves as their fitting curves using Gaussian functions. It can be seen that the results are similar with those in Fig. 2, with coincidence peaks and dips when the two photons arrive the fiber coupler simultaneously. The visibilities of the results of "$M_1$" and "$M_2$" are 78% and 81%, respectively, calculated by Eq. (2). It shows that the fidelities of $|\psi^-\rangle$ and $|\psi^+\rangle$ are 91%±3% and 88%±3%, respectively. They are a little smaller than the results in Fig. 2, showing the impact of fiber transmission on the BSM. Hence, the experiment results demonstrate the feasibility of the encoding/decoding processes. However, it is worth noting that the error rate of QSDC system based on this setup depends on the fringe visibilities shown in Fig. 3, which are still limited. It can be expected that further works are required to improve the visibilities, for example, by using narrower optical filters [22] in the photon pair source and temperature control on the fibers for quantum memories to reduce the polarization variation of photons in them.

**Discussion and conclusion**

The experiment results show that this experiment setup provides a proto-type of entanglement-based QSDC over optical fibers based on current optical communication technologies. In this system, the information is carried on polarization entangled photon pairs and decoded by BSM. Optical fibers are used as quantum channels for transmission and quantum memories at both sides. Comparing with quantum memories based on atomic ensembles and ions in crystals, optical fibers are more simple, reliable, and with high efficiency and fidelity if the required store time is within several tens microseconds, especially at telecom band. Hence, it can be expected that practical QSDC systems could be developed based on it by replacing all the wave plates by polarization modulators and adding electrical control to realize the protocol dynamically. Furthermore, in the experiment only two Bell states are used. The capacity could be improved by the complete hyper-entangled Bell-state analysis, in which sixteen Bell states can be completely distinguished, in principle [23].

Here, we theoretically evaluate the performance of QSDC systems developed by this scheme under typical parameters of current optical fiber components. In such a QSDC system, the polarization entangled photon pairs are generated by spontaneous four wave mixing in optical fibers. The generation rate is determined by the repetition rate of the pump pulses and the possibility of photon pair generation per pump pulse, which are denoted by $F_p$ and $p$, respectively. The photons "B" of the generated pairs are sent to Bob over optical fiber, the fiber length is denoted by $l_t$. Then half of them are selected by a fiber coupler and stored in the quantum memory at Bob's side, which is a piece of fiber with a length of $l_m$. The storage time of the quantum memories is $l_m/v$, in which $v$ is the light speed in optical fibers and about 200, 000 km/s. Finally, these photons are sent to the BSM system. On the other hand, the photons "A" of the generated pairs stay at Alice side. Half of them are selected by a fiber coupler and stored in the quantum memory at Alice side. The quantum memory is a also piece of optical fiber with the same length as that at Bob side ($l_m$). Then, the information is encoded on these photons by polarization modulation when they output from the quantum memory, the loss of the polarization modulator is denoted by $L_m$. These photons are sent to Bob by another piece of fiber with a length of $l_t$. Finally, at Bob's side these photons are also sent to the BSM system. The information is decoded by the BSM on the pairs. The collection efficiency of the BSM setup for each photon is denoted by $\eta_{BSM}$, which includes the efficiency of single photon detectors and the losses of the optical components such as the fiber coupler, polarization controllers and ODL. In the analysis, attenuations of the fibers for transmission and memory are the same, and denoted by $\alpha$.

The coincidence counting rate of the BSM is denoted by $R$. It can be expressed as

$$R = \frac{1}{4} F_p p e^{-2\alpha(l_t + l_m)} \eta_{BSM}^2 \tag{3}$$

According to the QSDC protocol, the storage time of the quantum memories should satisfy that

$$\frac{l_m}{v} \geq \frac{2l_t}{v} + t \tag{4}$$

Where, the term of $2l_t/v$ is the time that Alice sends photons to Bob over fiber, and Bob sends back the detection results of the photons for security test. $t$ is the time to record enough coincidence counts in the security test. If the requirement of the coincidence counts for the security test is $m$, $t$ can be expressed as

$$t = \frac{4m}{F_p p e^{-\alpha l_t} \eta_s^2} \tag{5}$$

where $\eta_s$ is the collection efficiency of the setup for the security test at each side, including the efficiency of the single photon detector and the losses of the polarization modulator and PBS. We assume that $\eta_s$ is the same at both sides. In the analysis, the data processing time for the security test is neglected.

According to Eq. (3), (4) and (5), we can calculate the maximum coincidence rate $R_{max}$ of the

BSM in the QSDC system under different transmission distances by setting the equal sign in Eq. (4). $R_{max}$ is a proper parameter to evaluate the performance of this entanglement-based QSDC system since it determines its capacity on information transmission. The main parameters used in the calculation are listed in Table 2, according to the current technologies of optical components and superconductor nanowire single photon detectors (SNSPDs). The efficiencies of detectors are set as 75%. Besides the efficiency of detectors, $\eta_{BSM}$ and $\eta_s$ also includes extra losses of -5 dB and -4 dB, respectively, for the optical components such as polarization controllers, PBSs, optical switchs, polarization modulator and ODL, and so on. The parameters for system optimization are photon pair generation possibility per pump pulse ($p$) and the required coincidence counts for the security test ($m$).

The calculation results are shown in Fig. 4. Fig. 4(a) is the relation between $R_{max}$ and the transmission distance $l_t$ under different photon pair generation possibility ($p$) per pump pulse. In the calculation, the coincidence number for the security test ($m$) is 1000. It can be seen that under a specific $p$, $R_{max}$ decreases with the transmission distance monotonously due to the loss of the transmission fiber. Higher $p$ is helpful to increase $R_{max}$, however, if $p$ is too large, the possibility of multi-pair generation will increase rapidly and impact the security of the QSDC. Fig. 4(b) is the calculation results under different coincidence counts $m$ required for the security test when $p$ is 0.03. It can be seen that higher requirement on $m$ means longer time is needed to take the security test, which increases the storage time at both sides and introduces larger losses of the quantum memories. Hence, the system performance could be improved by high efficiency security test which requires smaller coincidences ($m$). If $m$ is set to 1000, $R_{max}$ would be higher than 100 kHz under $l_t$ = 1km. If the transmission distance $l_t$ increases to 10km, $R_{max}$ would be higher than 3 kHz. If $m$ reduces to 100, $R_{max}$ would be higher than 100 kHz under $l_t$=10km. If the transmission distance $l_t$ increases to 25km, $R_{max}$ would be higher than 1.7 kHz. It can be seen that this fiber based QSDC system has the potential to realize a transmission rate close to security key rates of current commercial QKD systems (http://www.idquantique.com/quantum-safe-crypto/qkd-overview/), with the advantage that the QSDC system could transmit not only secure keys but also the information directly.

**Acknowledgements**

This work was supported by National Key R&D Program of China under Contracts No. 2017YFA0303700, the 973 Programs of China under Contracts No. 2013CB328700; the National Natural Science Foundation of China under Contracts No. 61575102, 11474168 and No. 61621064; the Tsinghua University Initiative Scientific Research Program

**Conflict of Interests**

The authors declare that they have no conflict of interests

**References**

[1] Bennett CH. Quantum cryptography using any two nonorthogonal states. *Phys Rev Lett* 1992; **68**: 3121–3124.

[2] Bennett CH, Brassard G. Quantum cryptography: public key distribution and coin tossing. In: *Proceedings of the IEEE International Conference on Computers, Systems and Signal Processing, Bangalore, India.* IEEE: New York, USA; 1984, pp 175–179.

[3] Ekert AK. Quantum cryptography based on Bell theorem. *Phys Rev Lett* 1991; **67**: 661.

[4] Marcikic I, de Riedmatten H, Tittel W, Zbinden H, Gisin N. Long-distance teleportation of qubits at telecommunication wavelengths. *Nature* 2008; **421**: 509.

[5] Sun QC *et al*. Quantum teleportation with independent sources and prior entanglement distribution over a network. *Nat. Photonics* 2016; **10**: 671.

[6] Takesue H *et al.* Quantum teleportation over 100 km of fiber using highly efficient superconducting nanowire single-photon detectors. *Optica* 2015; **2**: 832.

[7] Tittel W, Zbinden H, Gisin N. Experimental demonstration of quantum secret sharing. *Phys. Rev. A* 2001; **63**: 042301.

[8] Wang S *et al.* Experimental demonstration of a quantum key distribution without signal disturbance monitoring. *Nat. Photonics* 2015; **9**: 832.

[9] Takesue H, Sasaki T, Tamaki K, Koashi M. Experimental quantum key distribution without monitoring signal disturbance. *Nat. Photonics* 2015; **9**: 827.

[10] Korzh B, Lim CCW, Houlmann R, Gisin N, Li MJ, Nolan D, Sanguinetti B, Thew R, Zbinden H, Provably secure and practical quantum key distribution over 307 km of optical fibre. *Nat. Photonics* 2015; **9**: 163.

[11] Gisin N, Thew R. Quantum communication. *Nat. Photonics* 2007; **1**: 165.

[12] Kimble HJ, The quantum internet. *Nature* 2008; **453**, 1023.

[13] Long GL, Liu XS. Theoretically efficient high-capacity quantum-key distribution scheme. *Phys Rev A* 2002; 65: 032302.

[14] Deng FG, Long GL, Liu XS. Two-step quantum direct communication protocol using the Einstein-Podolsky–Rosen pair block. *Phys Rev A* 2003; 68: 042317.

[15] Deng FG, Long GL. Secure direct communication with a quantum one-time pad. *Phys Rev A* 2004; 69: 052319.

[16] Horodecki R, Horodecki P, Horodecki M, *et al*. Quantum entanglement. *Reviews of modern physics* 2009; 81: 865.

[17] Zhang W, Ding DS, Sheng YB, Zhou L, Shi BS, Guo GC. Quantum secure direct communication with quantum memory. *Phys Rev Lett* 2017; **118**: 220501.

[18] Hu J *et al*. Experimental quantum secure direct communication with single photons. *Light: Science & Application* 2016; **5**: e16144.

[19] James DFV, Kwiat PG, Munro WJ, White AG. Measurement of qubits. *Phys. Rev. A* 2001; **64**: 052312.

[20] Zhu F, Zhang W, Huang YD. Fiber based frequency-degenerate polarization entangled photon pair sources for information encoding. *Opt. Express* 2016; **24**: 25619-25628.

[21] Aspect A, Grangier P, Roger G, Experimental realization of Einstein-Podolsky-Rosen-Bohm gedankenexperiment - a new violation of Bell inequalities. *Phys Rev Lett* 1982; **49**: 91.

[22] Zukowski M, Zeilinger A, Weinfurter H. Entangling photons radiated by independent pulsed sources. *Ann. N.Y. Acad. Sci.* 1995; **755**: 91.

[23] Sheng YB, Deng FG, Long GL. Complete hyperentangled-Bell-state analysis for quantum communication. *Phys Rev A* 2010; **82**: 032318.

## Figure captions

Figure 1

The experimental setup. The source is a fiber-based frequency-degenerated polarization entangled photon pair source at telecom band. 50:50 fiber couplers at Alice's and Bob's parts are used as switches for the photons 'A' and 'B', respectively. The half wave plates, polarization beam splitters, single photon detectors and the TCSPC module in the setup (HWP-2, PBS-2 and SPD-2 at the part of Alice, HWP-1, PBS-1 and SPD-1 at the part of Bob) are used to realize the Bell's inequality test. DSFs at Alice and Bob are used as the quantum memories (QM-1 and QM-2). The two quarter wave plates (QWP-1 and QWP-2) are used to realize information encoding on the photons 'A'. The inset figure is the setup of the BSM system. The main part includes a 50:50 fiber coupler and two polarization beam splitters (PBS-5 and PBS-6). PC-1, PC-2 and PC-3 are polarization controllers, ODL is an optical delay line. PBS-3 and PBS-4 are also polarization beam splitters. SPD-3 and SPD-4 are two single photon detectors, TCSPC is the time-correlated single photon counting module.

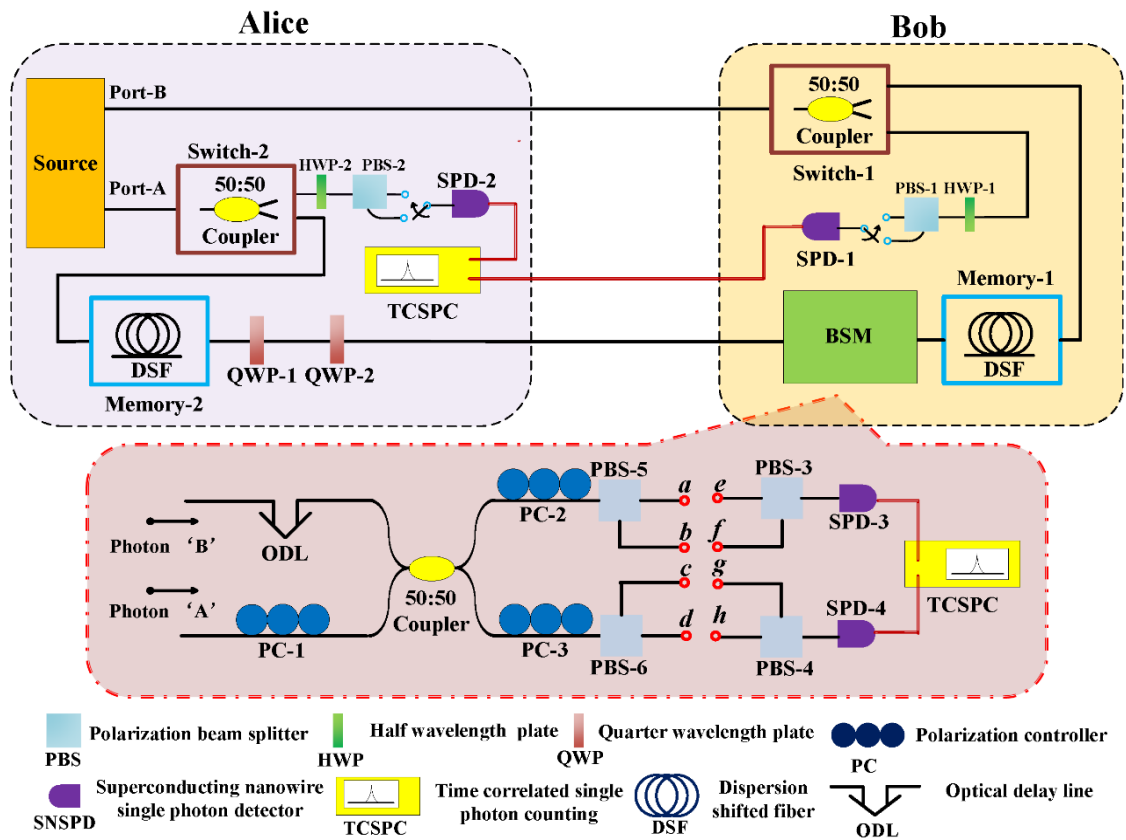

Figure 2

The results of BSM under different time delay without the fibers for transmission and quantum memories. (a) The coincidence counts under the measurement setting of "$M_1$". The hollow squares and circles are the coincidence counts when the photon pairs are in $|\psi^-\rangle$ and $|\psi^+\rangle$, respectively. The red and green curves are their fitting curves using Gaussian functions. (b) The coincidence counts under the measurement setting of "$M_2$". The hollow squares and circles are the coincidence counts when the photon pairs are in $|\psi^-\rangle$ and $|\psi^+\rangle$, respectively. The red and green curves are their fitting curves using Gaussian functions.

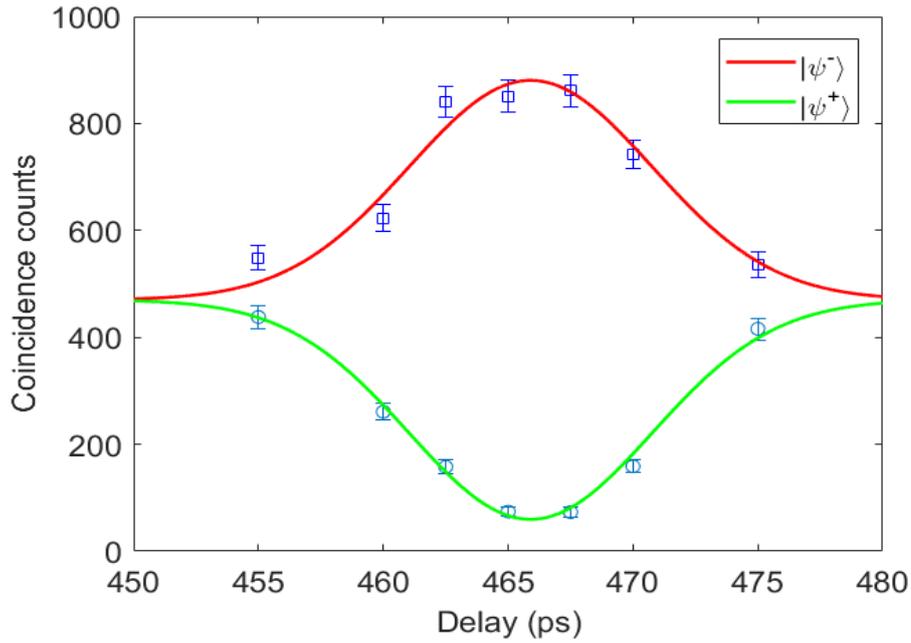

(a)

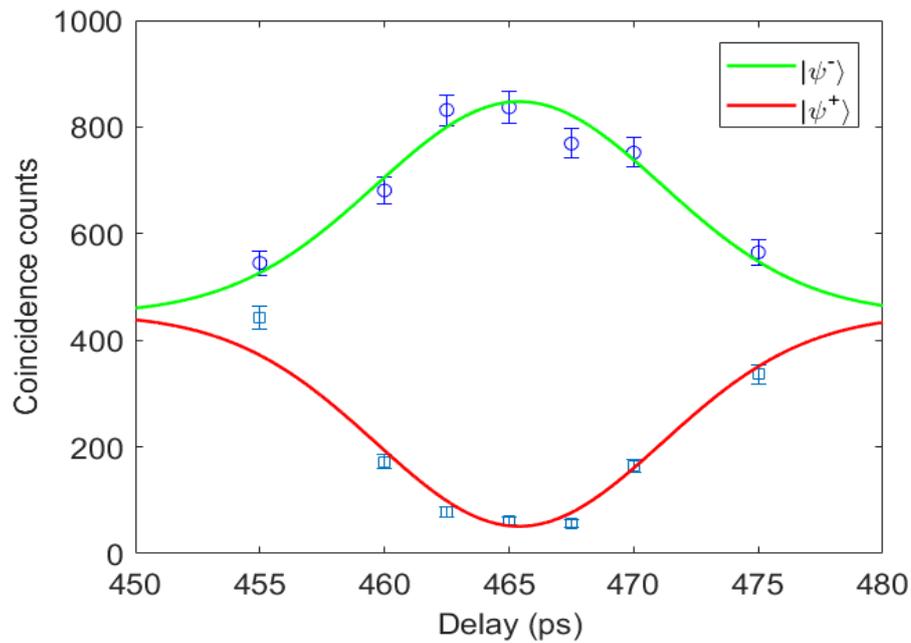

(b)

Figure 3

The results of the BSM under different time delay with the fibers for transmission and quantum memories. (a) The coincidence counts under the measurement setting of "$M_1$". The hollow squares and circles are the coincidence counts when the photon pairs are in $|\psi^-\rangle$ and $|\psi^+\rangle$, respectively. The red and green curves are their fitting curves using Gaussian functions. (b) The coincidence counts under the measurement setting of "$M_2$". The hollow squares and circles are the coincidence counts when the photon pairs are in $|\psi^-\rangle$ and $|\psi^+\rangle$, respectively. The red and green curves are their fitting curves using Gaussian functions.

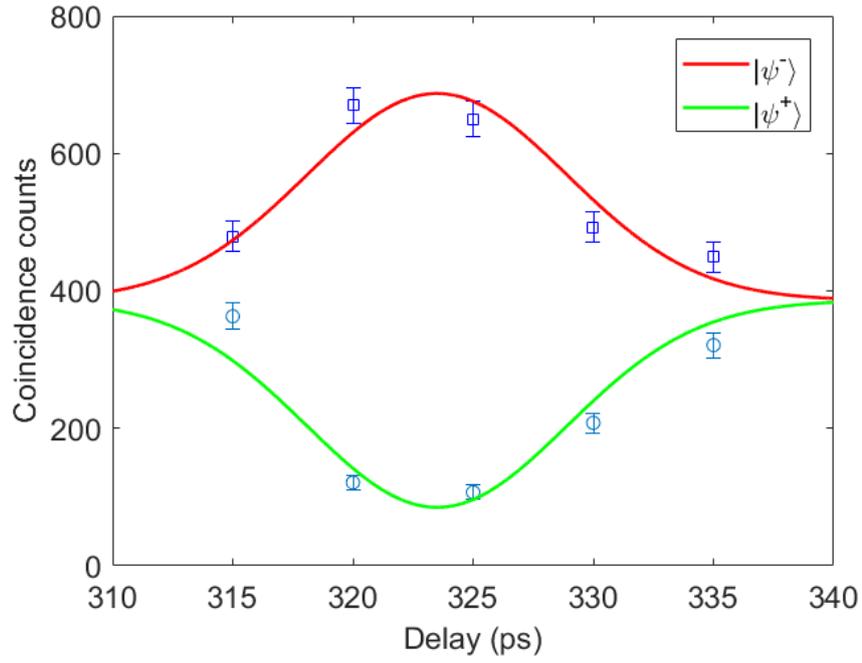

(a)

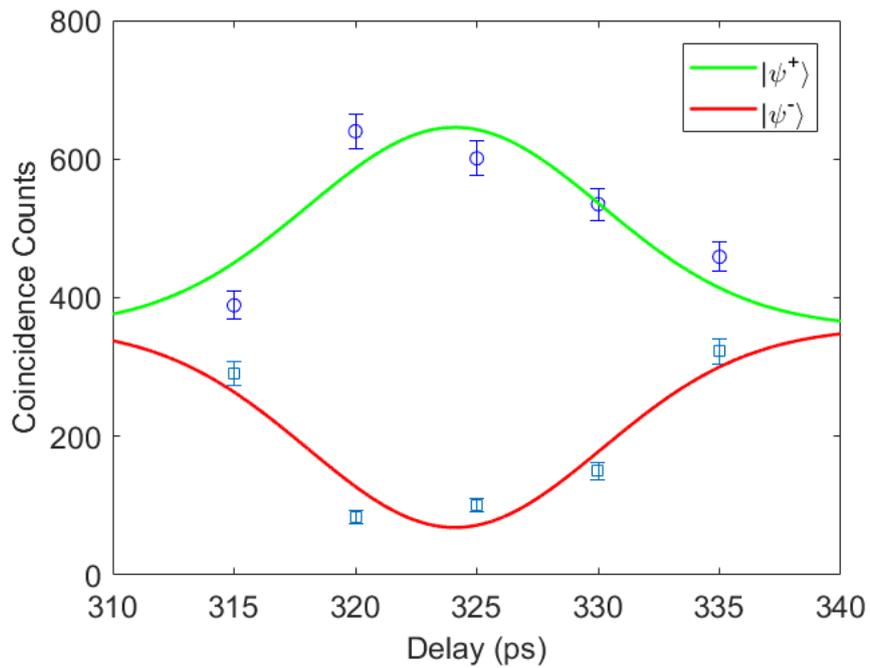

(b)

Figure 4

The performance evaluation of the entanglement-based QSDC system. (a) The relation between $R_{max}$ and $l_t$ when $p$ varies from 0.015 to 0.03 and $m=1000$. (b) The relation between $R_{max}$ and $l_t$ when $m$ is ranging from 100 to 2000 and $p=0.03$.

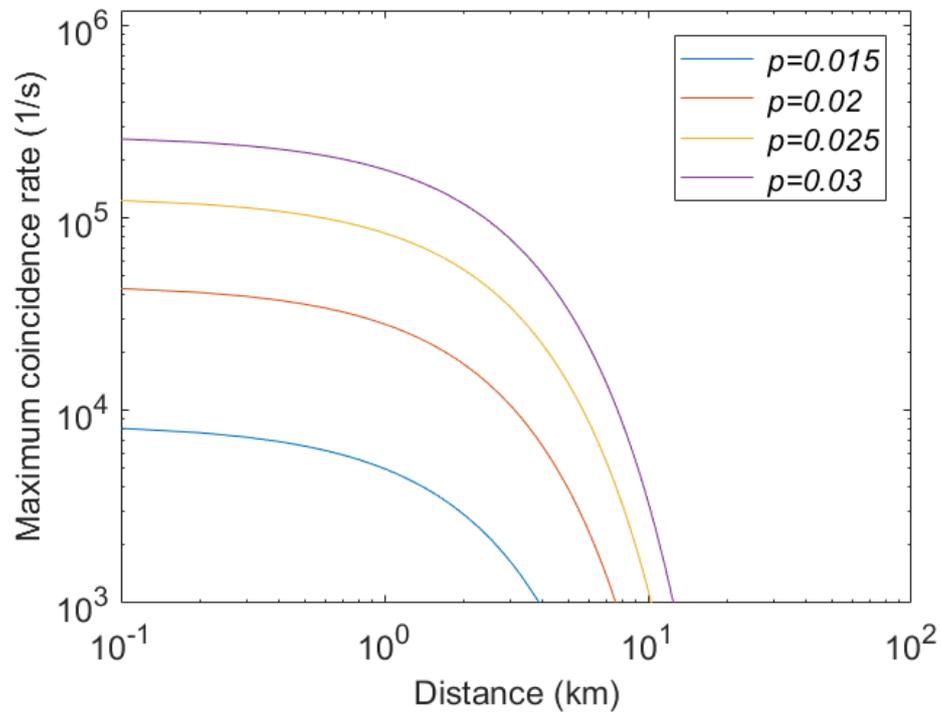

(a)

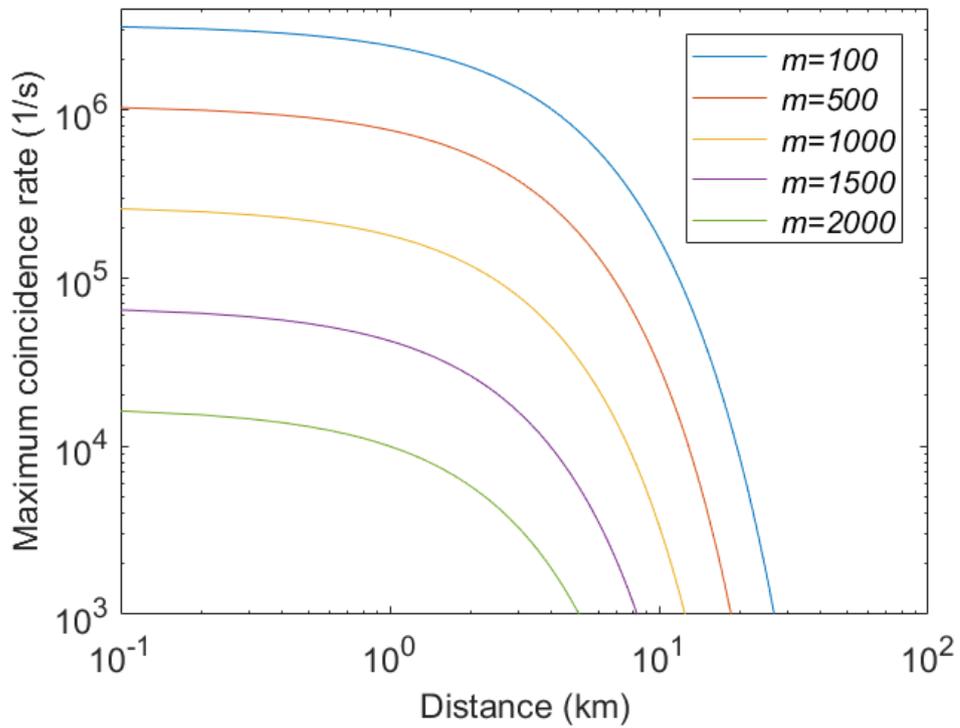

(b)

**Tables**

Table 1
Experimental results of the Bell's inequality test

| $\theta_1$ | $\theta_2$ | $N_{i,k}(\theta_1, \theta_2)$ | $N_{j,l}(\theta_1, \theta_2)$ | $N_{i,l}(\theta_1, \theta_2)$ | $N_{j,k}(\theta_1, \theta_2)$ |
|---|---|---|---|---|---|
| 0° | 11.25° | 420 | 378 | 101 | 139 |
| 22.5° | 11.25° | 433 | 390 | 89 | 122 |
| 22.5° | 33.75° | 460 | 421 | 58 | 101 |
| 0° | 33.75° | 112 | 75 | 403 | 449 |

Table 2
Main parameters for the performance evaluation of the QSDC system.

| Attenuation of optical fibers | $\alpha$ | 0.2 dB/km |
|---|---|---|
| Repetition rate of pulsed pump light | $F_p$ | 10 GHz |
| Collection efficiency of the BSM system | $\eta_{BSM}$ | 23.7% |
| Collection efficiency of the setup for the security test | $\eta_s$ | 29.9% |